\documentclass{elsart}
\usepackage{epsfig}
\usepackage{amssymb}

\begin{document}
\newcommand{\EGLOB}{E_{\rm glob}}
\newcommand{\ELOC}{E_{\rm loc}}
\newcommand{\ERANDGLOB}{E_{\rm glob}^{\rm random}}
\newcommand{\ERANDLOC}{E_{\rm loc}^{\rm random}}
\newcommand{\be}{\begin{equation}}
\newcommand{\ee}{\end{equation}}
\newcommand{\bea}{\begin{eqnarray}}
\newcommand{\n}{\nonumber\\}
\newcommand{\eea}{\end{eqnarray}}

\begin{frontmatter}
\title{Efficiency of Scale-Free Networks:
\\
Error and Attack Tolerance }

\author[label1]
{Paolo Crucitti},
\author[label2]
{Vito Latora}, 
\author[label3,label4]
{Massimo Marchiori}, and
\author[label2]
{Andrea Rapisarda} 

\address[label1] {Scuola Superiore di Catania,
Via S. Paolo 73, 95123 Catania, Italy}

\address[label2] {Dipartimento di Fisica e Astronomia,
Universit\`a di Catania,\\
and INFN sezione di Catania, Corso Italia 57, 95129 Catania, Italy}

\address[label3] {W3C and Lab. for Computer Science,
Massachusetts Institute of Technology, USA}

\address[label4]
{Dipartimento di Informatica, Universit\`a di Venezia, Italy}

\begin{abstract}
The concept of network efficiency, recently proposed
to characterize the properties of small-world networks, is here
used to study the effects of errors and attacks on scale-free 
networks. 
Two different kinds of scale-free networks, i.e. networks with power 
law P(k), are considered: 1) scale-free networks with no local clustering 
produced by the Barabasi-Albert
model and 2) scale-free networks with high clustering properties
as in the model by Klemm and Egu\'{\i}luz, 
and their properties are compared to the properties of random 
graphs (exponential graphs). 
By using as mathematical measures the global and the local
efficiency we investigate the effects of errors and attacks 
both on the global and the local properties of
the network. We show that the global efficiency is a better measure 
than the characteristic path length to describe the response 
of complex networks to external factors. 
We find that, at variance with random graphs, 
scale-free networks display, both on a  
global and on a local scale, a high degree of error tolerance and 
an extreme vulnerability to attacks. In fact, the global and the local 
efficiency are unaffected by the failure of some randomly 
chosen nodes, though they are extremely sensititive 
to the removal of the few nodes which play a crucial 
role in maintaining the network's connectivity.
\end{abstract}


\end{frontmatter}

\section{Introduction}
\label{intro}
The study of the structural properties of the underlying network 
can be very important to understand the functions of a complex 
system \cite{yaneer}. 
For instance the architecture of a computer network 
is the first critical issue to take into account 
when we want to design an efficient communication system. 
Similarly, the efficiency of the communication and of the navigation 
over the Net is strongly related to the topological properties of 
the Internet and of the World Wide Web. 
The connectivity structure of a population (the set 
of social contacts) affects the way ideas are diffused, but  
also the spreading of epidemics over the network. 
Only very recently the increasing accessibility of databases 
of real networks on one side, 
and the availability of powerful computers on the other side, 
have made possible a series of empirical studies on the 
properties of biological, technological and social networks.  
The results obtained have shown that, in most  cases,
real networks {\it are very different from random and regular
networks}, and {\it display some common properties as 
high efficiency and high degree of robustness}.
The literature on complex networks has followed an exponential 
growth in the last few years; a comprehensive review can be found 
in Refs.\cite{revstrogatz,revbarabasi,newman1}. 
In the following we enumerate some of the results appeared in
the recent literature that are important in order to understand the
purpose of this paper:

\begin{enumerate}

\item  
In ref. \cite{watts}, Watts and Strogatz
have shown that the connection topology of some real networks 
is neither completely regular nor completely random. 
These networks, named {\it small-world networks} \cite{milgram}, 
exhibit in fact {\it high clustering coefficient}, like regular lattices, 
and {\it small average distance} between two generic points 
(small characteristic path length), 
like random graphs. Watts and Strogatz have also proposed 
a simple model (the WS model) to construct networks with small-world 
properties (i.e. networks with high clustering and small average 
distance), by rewiring few edges of a regular 
lattice. 

\item 
In ref.\cite{lm2} two of us have
introduced the concept of {\em efficiency\/} of a
network, which measures how efficiently the information
is exchanged over the network.
By using the efficiency as a new measure to characterize the network,
it has been showed that small-worlds are systems that are
both globally and locally efficient. Moreover the description of
a network in terms of its efficiency extends the
small-world analysis also to unconnected networks and
to real systems that are better represented as
weighted networks \cite{lm3,lm4,lm1}.

\item  
Small average distance and high clustering are not all the common 
features of complex networks. 
Barabasi and collaborators have studied $P(k)$, 
the degree distribution of a network, and found that 
many large networks (the World Wide Web, Internet,
metabolic networks and protein networks) are 
{\it scale-free}, i.e. have a {\it power-law degree 
distribution} $P(k) \sim k^{-\gamma}$ 
\cite{barabasi0,barabasi1,barabasi2,barabasi3att,barabasi4met,barabasi5prot}.
Neither random graph theory \cite{bollobas},
nor the WS model  to construct networks with the
small-world properties \cite{watts}
can reproduce this feature: in fact both give 
$P(k)$ peaked around the average value of $k$. 
In ref. \cite{barabasi1} Barabasi and Albert 
have proposed a simple model (the BA model)
to construct a scale-free topology by
modeling the dynamical growth of the network:
some ad hoc assumptions in the network dynamics result
in a network with the correct scale-free features,
i.e. with a power-law degree distribution
$P(k) \sim k^{-3}$.

Moreover in ref.\cite{barabasi3att} the authors have shown
that scale-free networks,  at variance with  random networks,
display a high degree of error tolerance. That is
the ability of their nodes to communicate is unaffected by
the failure of some randomly chosen nodes.
However, error tolerance comes at a high price
in that scale-free networks are extremely vulnerable to attacks, i.e. to
the removal of a few nodes which  play a crucial  role in maintaining
the network's connectivity.
Such {\it error tolerance} and {\it attack vulnerability} typical 
of scale-free networks have also been found in real networks 
\cite{barabasi3att}.

\item 
The BA scale-free model produces networks with a power law 
connectivity distribution, but not with small-world properties. 
In fact the BA scale-free networks have 
small average distance between two generic points,
the first property of a small-world network, 
while they lack of high clustering, 
the other property of a small-world network. 
More recently  Klemm and Egu\'{\i}luz \cite{klemm2} have
proposed an alternative model (the KE model) 
to construct networks where scale-free degree distributions 
coexist with small average distances
and with strong clustering. Therefore,
the KE model reproduces,
at the same time, the two distinct features present in real
networks: power law degree distribution and the
small-world behavior.
\end{enumerate}

In this paper we use the concept of global and local efficiency
to characterize the properties of scale-free networks
(i.e. networks with power law degree distributions), and
to study their error and attack tolerance.
We consider both scale-free networks with no clustering (the BA model), 
and scale-free networks with high clustering properties 
(the KE model). 
We analyze the effect of errors and attacks not only on the global 
properties of the network (as done in ref.\cite{barabasi3att} 
by using as a measure the average distance between two points) 
but also on the local properties of the network.
Moreover we compare the results obtained in terms of global 
and local efficiency of the network 
with the results in terms of average distance and clustering 
coefficient. The three innovative point of our paper are: 

\begin{itemize}
\item 
The use of the efficiency measure to characterize scale-free networks.
This allows to avoid problems due to the divergence of the average 
distance. 

\item  
The parallel study of scale-free networks
with no clustering, and scale-free networks with high clustering.

\item  
The study of the effect of errors and attacks not only on the global
propertied, but also on the local properties of the network.
\end{itemize}

The paper is organized as follows. In Section \ref{swn} we define the 
variable efficiency and we illustrate how the small-world 
behavior can be expressed in terms of the local 
and the global efficiency of the network. 
In Section \ref{sfn} we discuss the relevance and the properties 
of scale-free networks, and we illustrate the BA model and the 
KE model. 
In Section \ref{esfn}, the central part of the paper, 
we investigate the effects of errors and attacks both on 
the global and on the local properties of scale-free networks. 
We show that the efficiency is a better measure 
than the characteristic path length to describe the global 
properties of complex networks, especially when a large number 
of nodes is removed. 
The local properties of the scale-free networks are equally well  
described by the local efficiency or by the clustering coefficient. 
By considering both BA and KE scale-free networks, we show 
that scale-free networks are systems resistent to errors 
but vulnerable to attacks both at a global and at a 
local level. In Section \ref{conclusions} we draw the conclusions.

\section{Small-World behavior and Efficiency of a Network}
\label{swn}
In their seminal paper Watts and Strogatz have shown that
the connection topology of some real (biological, social
and technological) networks is neither completely regular
nor completely random \cite{watts}. Watts and Strogatz have named
these networks, that are somehow in between regular and
random networks, {\it small-worlds},
in analogy with the small-world phenomenon,
empirically observed in social systems more than 
30 years ago \cite{milgram}. 
The mathematical characterization of the small-world
behavior is based on the evaluation of two quantities,  
the characteristic path length $L$, measuring the 
typical separation between two generic nodes in the network 
and the clustering coefficient $C$, 
measuring the average cliquishness of a node.
Small-world networks are in fact highly clustered, like regular 
lattices, yet having small characteristic path lengths, 
like random graphs. Let us give some useful mathematical 
formalism. 
A generic unweighted (or relational) network \cite{lm4} 
is represented by a graph $\bf G$ with $N$ 
vertices (nodes) and $K$ edges (arcs, links or connections). 
Such a graph is described by the so-called  
adjacency  matrix $\{a_{ij}\}$ (also called connection matrix).
This is a $N \cdot N$ symmetric matrix,
whose entry $a_{ij}$ is $1$ if there is an
edge joining vertex $i$ to vertex $j$, and $0$ otherwise.
An important quantity of graph $\bf G$, which will be used in the
following of this paper, is the degree of a generic vertex $i$,
i.e. the number $k_i$ of edges incident with vertex $i$,
the number of neighbours of $i$. We have
$K= \sum_i k_i /2$ because each link is counted twice, 
and the average value of $k_i$ is $ <k> = 2 K/N$. 
To define $L$ we need first to construct the shortest path length
$d_{ij}$ between two vertices (known in social networks studies 
as the number of degrees of separation
\cite{milgram}), measured as  the miminum number
of edges traversed to get from a vertex $i$ to another vertex $j$.
By definition $d_{ij} \ge 1$ with  $d_{ij}=1$ if there exists a direct
edge between $i$ and $j$.
The characteristic path length $L$ of graph $\bf G$
is defined as the average of the shortest path lengths
between two generic vertices:
\be
L({\bf G}) = \frac {1}{N(N-1)} \sum_{i\neq j \in {\bf G} } d_{ij}
\label{l}
\ee
Of course this definition is valid only if {\bf G}  is totally  connected,  which means that
 there must exist at least a path connecting any couple of vertices
with a finite number of steps.
Otherwise, when from $i^*$ we can not reach $j^*$ then
$d_{i^*j^*}=+ \infty $ and consequentely $L$ as given in eq.(\ref{l}),being divergent, is an ill-defined quantity.
When studying how the properties of a network
are affected by the removal of nodes, one often incurrs  in
non-connected networks. In such cases the alternative formalism
in terms of efficiency here proposed is much more powerful, 
as will be clarified in the following.
\\
The second measure, the clustering coefficient $C$, 
is a local quantity of ${\bf G}$ defined as follows. 
For any node $i$ we consider $\bf G_i$, the subgraph of neighbors of $i$. 
That is once eliminated $i$ we study how the nodes 
previously connected to $i$ remain still connected between each other.
If the node $i$ has $k_i$ neighbors, then $\bf G_i$ has
$k_i$ nodes and at most $k_i(k_i-1)/2$ edges.
$C_i$ is the fraction
of these edges that actually exist, and $C$ is the average value
of $C_i$ all over the network:
\be
C({\bf G}) = \frac {1}{N} \sum_{i \in {\bf G} } C_i ~~~~~~~~
C_i = {  \mbox{  \# of edges in  $\bf G_i$ }
        \over
        \mbox{  $k_i(k_i-1)/2$ ~~}
     }
\ee
To illustrate the onset of the small-world, Watts and Strogatz have
proposed a one-parameter model (the WS model) to construct
a class of unweighted graphs which interpolates between a
regular lattice and a random graph.
The edges of a regular lattice are rewired with a probability
$p$. As the rewiring probability $p$ increases, the network
becomes increasingly disordered and for $p=1$ a random graph
is obtained. Although in the two limiting cases
large $C$ is associated to large $L$ ($p=0$)
and viceversa small $C$ to small $L$ ($p=1$),
there is an intermediate regime where the
network is a small-world:  highly clustered like a regular lattice
and with small characteristic path lengths like a
random graph. In fact only a  few rewired edges
($0 < p \ll 1$) are sufficient to produce
a rapid drop in $L$, while $C$ is not affected and
remains equal to the value for the regular lattice \cite{watts}.
By means of this mathematical formalism based on the evaluation 
of $L$ and $C$, Watts and Strogatz have found three 
examples of small-world behavior in real networks:  
1) the collaboration graph
of actors in feature films from Ref.\cite{actors}, 
as an example of a social system; 2)  
the neural network of a nematode, the  C.~elegans \cite{verme}
as an example of a biological network; 3) finally an example of 
a technological network, the electric power grid of the western
United States.

An alternative definition of the small-world behavior has been
proposed more recently by two of us in ref.\cite{lm2,lm4}
and is based on the definition of the {\em efficiency\/} of a
network. Instead of $L$ and $C$ the network is characterized in
terms of how efficiently it propagates information on a global and on
a local scale, respectively.
To define the efficiency  of $\bf G$ let us
suppose that every node sends information along
the network, through its edges. We assume that the
efficiency $\epsilon_{ij}$ in the communication between
node $i$ and $j$ is inversely proportional
to the shortest distance:  $\epsilon_{ij} = 1/d_{ij} ~\forall i,j$.
With this definition, when there is no path in the graph
between $i$ and $j$, $d_{ij}=+\infty$ and consistently
$\epsilon_{ij}=0$.
The {\it global efficiency} of the graph $\bf G$
can be defined as:
\begin{equation}
\label{globaleff}
\EGLOB({\bf G})=
\frac{ {{\sum_{{i \ne j\in {\bf G}}}} ~\epsilon_{ij}}  } {N(N-1)}
          = \frac{1}{N(N-1)}
{\sum_{{i \ne j\in {\bf G}}} \frac{1}{d_{ij}}}
\end{equation}
and the {\it local efficiency}, in analogy with $C$, can
be defined as the average efficiency of local subgraphs:
\begin{equation}
\label{localeff}
\ELOC({\bf G})   = \frac{1}{N} \sum_{i \in {\bf G}} ~ {E(\bf {G_i})}
~~~~~~~
{E(\bf {G_i})}
          = \frac{1}{k_i(k_i-1)}
{\sum_{{l \ne m\in {{\bf G_i}}}} \frac{1}{d^{\prime}_{lm}}}
\end{equation}
\noindent
where $\bf G_i$, as previously defined, is the subgraph of
the neighbours of $i$, which is made by $k_i$
nodes and at most ${k_i(k_i-1)}/2$ edges.
It is important to notice that the quantities $\{d^{\prime}_{lm}\}$
are the shortest distances between nodes $l$ and $m$
calculated on the graph $\bf G_i$.
The two definitions we have given have the important property that
both the global and local efficiency are already normalized, that is:
$0 \le \EGLOB({\bf G}) \le 1$ and $0 \le \ELOC({\bf G}) \le 1$ \cite{normalization}.
The maximum value of the efficiency $\EGLOB({\bf G})= 1$
and $\ELOC({\bf G}) = 1$ are obtained in the
ideal case of a completely connected graph, i.e. in the case
in which the graph $\bf G$ has all the $N(N-1)/2$ possible edges
and $d_{ij} =1~\forall i,j$.
In the efficiency-based formalism a small-world results as a
system with high $\EGLOB$ (corresponding to low $L$) and high
$\ELOC$ (corresponding to high clustering $C$),
i.e. a network extremely efficient in exchanging information
both on a global and on a local scale.
Moreover the description of a network in terms of its efficiency
extends the small-world analysis also to unconnected networks and, more
important, with only a few modifications, to weighted networks.
A weighted network is a case in which there is a weight
associated to each of the edges. Such a network needs two
matrices to be described: the usual adjacency matrix
$\{a_{ij}\}$ telling about the existence or not existence
of a link (and whose entry $a_{ij}$, as for the unweighted case, is
$1$ when there is an edge joining $i$ to $j$, and $0$ otherwise) and
and a second matrix, the matrix of the weights associated
to each link.
All the details of the applications of the efficiency-based formalism
to study real weighted networks,
e.g. the Boston subway transportation system,
can be found in \cite{lm2,lm3,lm4}.
In this paper we focus instead on the simpler case of
unweighted networks: we are in fact interested in the use of
the efficiency formalism to describe in quantitative terms the
global and the local properties of scale-free networks,
and to study how these properties are affected by
the random removal of nodes or by attacks.
%
\begin{figure}
\label{}
\begin{center}
\epsfig{figure=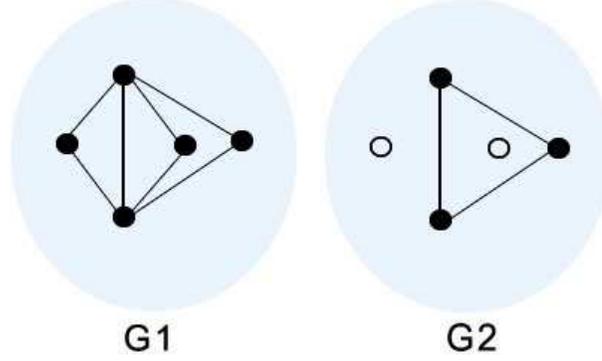,width=8truecm,angle=0}
\end{center}
\caption{ The connectivity properties of two graphs G1 and G2, 
both with N=5 nodes, are compared. 
Differently from the efficiency $\EGLOB$, the characteristic 
path length $L$ is not a representative measure when 
the graph is unconnected. 
At the local level, $C$ is a good approximation of $\ELOC$.
}  
\label{uncon}
\end{figure}
%
A simple example will be very useful to illustrate the comparison between 
$\EGLOB$, $\ELOC$ and $L$, $C$, and to explain why the efficiency in many 
cases works better than $L$ and $C$, even for unweighted networks. 
In particular the differences between the description in term of 
$\EGLOB$ and the description in terms of $L$ are evident 
when the network is unconnected. 
Fig.~\ref{uncon} is an example of the problems associated 
to the calculations of $L$ when the graph is unconnected. 
We consider 2 graphs 
$\bf G1$ and $\bf G2$, both having the same number of nodes $N=5$, 
but different number of edges. 
By using the definition (\ref{l}) we obtain 
$L_1=13/10$ for the graph on the left hand side and 
$L_2=\infty$ for the graph on the right hand side.  
An alternative possibility to avoid the divergence of $L_2$ 
is to limit the use of definition (\ref{l}) only to a part of 
the graph, the main connected component \cite{wattsactors} 
of $\bf G2$, which is made of 3 nodes. 
In this way we get $L_2=1$ and the final information we extract 
from the analysis of the characteristic path length is that 
graph $\bf G2$ has better structural properties than graph $\bf G1$, 
since $L_2<L_1$. 
This is of course wrong because $\bf G1$ is certainly much better 
connected than  $\bf G2$, and the misleading information comes from 
the fact that in the second graph we had to remove two 
nodes from the analysis.  
By studying instead the efficiency of the two graphs 
we are allowed to take into account also 
the nodes not connected to the main connected component: 
we get ${(\EGLOB)}_1=17/20$ and  ${(\EGLOB)}_2=3/10$, 
in perfect agreement with the fact that $\bf G1$ has a much better  
connectivity ($17/20$ the efficiency of the completely connected 
graph) than  $\bf G2$. 
\\
On the other side an evaluation of the local clustering of the 
two graphs gives: $C_1=4/5$, $C_2=3/5$, and an evaluation 
of the local efficiency gives:  ${(\ELOC)}_1=9/10$, 
${(\ELOC)}_2=3/5$. This indicates that the first graph has also 
better local properties than the second one. Moreover the variable 
$C$ is a good approximation of the local efficiency $\ELOC$
(this is in general true 
when the subgraphs $\bf G_i$ of a generic node 
$i$ are composed by small graphs \cite{lm4}).

\section{Scale-Free Networks}
\label{sfn}
An important information to characterize a graph $\bf G$, as 
previously mentioned, is the degree of a generic vertex $i$, 
i.e. the number $k_i$ of edges incident with vertex $i$, 
the number of neighbours of $i$.  
Barabasi and collaborators focussed their attention 
on $P(k)$, the degree distribution of a network, and 
showed that many real large networks, as the World Wide Web, 
the Internet, metabolic and protein networks,  
are {\it scale-free}, that is, their degree distribution 
follows a power-law for large $k$ 
\cite{barabasi0,barabasi1,barabasi2,barabasi3att,barabasi4met,barabasi5prot}. 
Also some social systems of interest for the spreading of 
sexually trasmitted diseases \cite{sta,ves}, 
and the connectivity network of atomic clusters' systems \cite{doye} 
display a similar behavior. 
Neither random graphs \cite{bollobas}, nor small-world networks 
constructed according to the WS model, have a power-law degree distribution 
$P(k)$ like the one observed in real large networks.  
In fact for a random graph $P(k)$ is described by a Poisson 
distribution $P(k) = {<k>^k}/k! ~e^{-<k>}$, a curve peaked 
at $k=<k>$ and exponentially decaying for large $k$, in 
contrast to the power-law decay of scale-free graph. 
This is the reason why random graphs are sometimes referred in the literature 
as exponential graphs \cite{barabasi3att}. 
Also in the case of the WS small-world model $P(k)$ is strongly 
peaked around the average value of $k$ (since it is very close 
to the $P(k)$ of regular graphs). 
Furthermore, even for those real networks for which $P(k)$ 
is not clearly a power law for all values of $k$, and has for  
instance an exponential cut off for very large $k$, the degree 
distribution significantly deviates from the Poisson expected for  
random graphs \cite{pnas}. 
At this point two natural questions come up to the mind: 
1) What is the mechanism responsible for the emergence 
of a scale-free structure in such a huge number of real networks ?   
2) What are the main properties of a scale-free topology, and 
why is it privileged with respect to the other topologies ? 

An answer to the first question and a concrete algorithm to 
construct a scale-free network has been proposed by Barabasi 
and collaborators. In Refs.\cite{barabasi1,barabasi2}  
the authors argue that the scale-free nature of 
real networks is rooted in two generic mechanisms 
occurring  in many real networks. 
First of all most real-world networks describe open systems 
which {\it grow} by the continuous addition of new nodes: as an 
example the WWW grows exponentially in time by the addition of 
new web pages, or the research literature constantly grows 
by the publication of new papers.
Moreover most real networks exhibit {\it preferential attachment}, 
that is, the likelihood of connecting to a node 
depends on the node's degree. A  webpage will most likely include 
hyperlinks to popular documents which have  already a high degree, 
because such highly connected documents are easier to find.   
A new manuscript will most  likely cite 
a well-known one increasing furthermore its high number of citations. 
Growth and preferential attachment are the two sufficient ingredients 
 to produce a scale-free 
network. The Barabasi-Albert (BA) model proposed 
in \cite{barabasi1,barabasi2} is a simple way to generate 
 a network with a power-law degree distribution 
$P(k) \sim k^{-\gamma}$, and with $\gamma=3$. 
On the contrary, neither of the two ingredients is present 
in the small-world model discussed in Section \ref{swn}, 
that assumes instead a fixed number $N$ of vertices and a 
probability that two nodes are connected (or their connection 
is rewired) independent of the nodes' degree. 

Concerning the second question, the authors of ref.\cite{barabasi3att} 
have studied the response of scale-free networks  
to errors and to attacks. 
By error and attack they indicate, respectively,  
the removal of randomly chosen nodes, and the removal 
of the most connected nodes. 
In particular they study the change of the characteristic 
path length $L$ when a small fraction of the nodes is eliminated: 
in fact the removal of a node in general increases the distance 
between the remaining nodes, because it can eliminate paths 
contributing to the connectivity of the system. 
Differently from random networks, the scale-free networks
display a high degree of error tolerance, i.e.  
the ability of their nodes to communicate is unaffected by 
the failure of some randomly chosen nodes.  
On the contrary these networks are extremely vulnerable to attacks, 
i.e. the removal of a few nodes that play a vital role in maintaining 
the network's connectivity. 
In practice the presence of the scale-free topology in 
so many real cases \cite{barabasi0,barabasi4met,barabasi5prot,sta} 
can be attributed to the need to construct systems with 
a high degree of tolerance against errors. 
Though the error tolerance comes at a high price 
in that the scale-free networks are extremely vulnerable 
to attacks. 
The response of scale-free networks to the removal 
of nodes is also one of the main points of our paper. 
In fact, in Section \ref{esfn} we will extend the analysis 
of ref.\cite{barabasi3att}, 
that was only based on the quantity $L$, 
to both the global and local properties of the network. 
In order to characterize the local properties of a graph 
we will use either $C$ and $\ELOC$. For the global properties 
we will see that $\EGLOB$ is better than $L$ especially when 
a large number of nodes are removed. 

The BA scale-free model reproduces the 
power-law connectivity distribution, but not the 
small-world effect. In fact it produces  networks  
with small average distance between two generic nodes, 
like  a small-world network, but lacks high 
clustering, which is typical of    a small-world network.
On the contrary, most large real networks with power-law 
connectivity distribution, shows also a high clustering coefficient. 
As an example the values of $C$ obtained from the two  
databases of Internet and of the World Wide Web studied, 
are orders of magnitude larger than the clustering coefficients 
for the correspective random graphs \cite{revbarabasi}. 
In order to overcome this problem Klemm and Egu\'{\i}luz \cite{klemm2} 
have recently proposed an alternative model, 
the KE model, which produces 
networks with scale-free degree distributions, 
small average distances and with strong clustering. 
With a minimal amount of changes to the BA model, the KE model 
reproduces, at the same time, 
the two distinct features of real networks: 
power-law degree distribution and small-world effect. 
We do not go into the details of the KE model now. 
Since the subject of this paper is the study of the 
properties of scale-free networks, in the next section  
we will discuss how to construct scale-free networks 
with the BA model, and scale-free networks 
with high clustering by means of  the KE model.

\section{Efficiency in Scale-Free Networks}
\label{esfn}
We are finally ready to study how the efficiency of a network 
with scale-free topology is affected by the removal of some 
of its nodes. 
We will make use of the measures defined in formula 
\ref{globaleff} and in \ref{localeff}, and compare 
the results with the ones obtained in terms of $L$ and $C$. 
The first step is the construction of a scale-free network: 
for this purpose we consider both the BA model and the KE model.

\subsection{Barabasi-Albert (BA) scale-free networks}
\label{basfn}
First we construct the scale-free network following 
the Barabasi-Albert (BA) model \cite{barabasi1,barabasi2}. 
As previously mentioned 
the two ingredients of the BA model are growth and preferential 
attachment. In fact the algorithm \cite{barabasi1}
is based on the iteration of the following two steps: 

(1) {\it Addition of nodes}:
Starting with a small number ($m_0$) of nodes, 
at every timestep a new node with $m$($\leq m_0$) edges 
is added. The edges link the new node to $m$ different nodes 
already present in the system. 

(2) {\it Preferential attachment of new edges}:
When choosing the nodes to which the new node connects, the 
probability $\Pi$ that the new node will be connected to node 
$i$ is assumed to depend on the degree $k_i$ of node $i$, according 
to:    
\begin{equation}
\label{pa}
\Pi(k_i)=\frac{k_i}{\sum_j k_j} 
\end{equation}
After $t$ timesteps the algorithm produces a network 
with $N=t+m_0$ nodes and $mt$ edges. 
The analytical solution of the BA model in the mean field 
approximation predicts a degree distribution 
$ P(k)= \frac {2m^{2}t} {m_0+t} k^{-3}$, 
This function asymptotically converges for $t\rightarrow \infty$ 
to a time-independent 
degree distribution $P(k)\sim 2m^{2} k^{-\gamma}$, 
i.e. to a power law with 
an exponent $\gamma=3$. It is interesting to notice that 
$\gamma$ does not 
depend neither on $m$ nor on the size $N=m_0+t$ of the network.
\\
The mean field predictions are confirmed by other 
analytical approaches (master equation \cite{doro} and 
rate equation \cite{ley}) and by numerical simulations.  
Both the two ingredients, growth and preferential attachment,  
are necessary in the BA model for the emergence of the 
power-law scaling.
Barabasi et al. have in fact checked that a model with 
 growth  but no preferential attachment gives for $t\rightarrow \infty$ an 
exponential degree distribution. On the other hand, 
a model with preferential attachment but no growth  predicts that 
the degree distribution becomes a Gaussian around 
its mean value. 
%
\begin{figure}
\begin{center}
\epsfig{figure=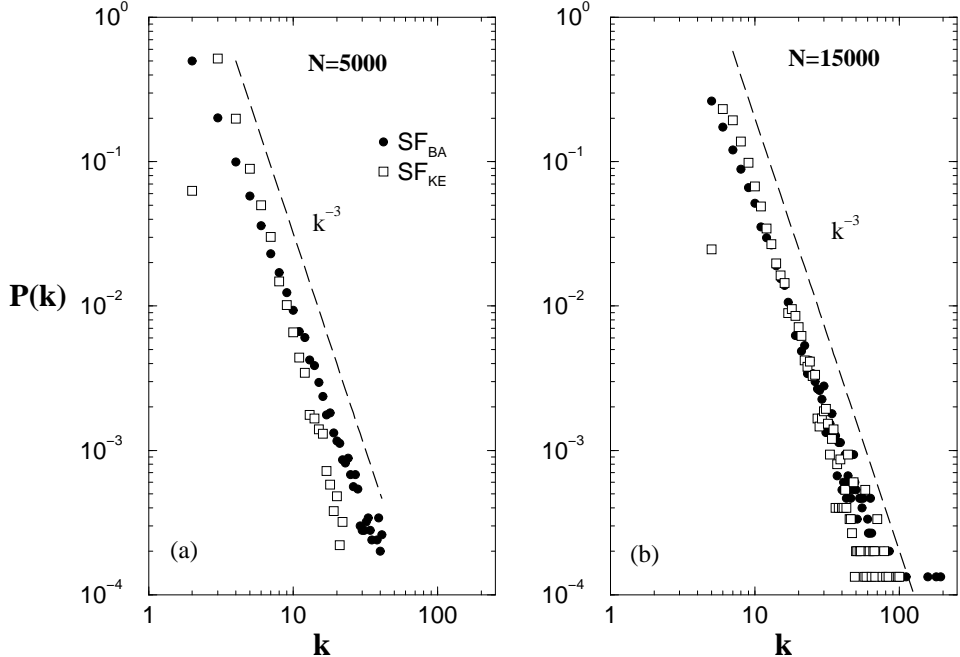,width=9truecm,angle=-90}
\end{center}
\caption{Degree distribution for the BA scale-free model (indicated 
as $SF_{BA}$ with full circles) 
and for the KE scale-free model (indicated 
as $SF_{KE}$ with open squares). 
Two system sizes are considered $N=5000$, $K=10000$ in (a), 
and $N=15000$, $K=75000$ in (b). 
For $N=5000$ the results reported are obtained as 
averages over 10 different realizations. While in the case 
$N=15000$ only one realization is considered.  
The dashed line is $P(k) \sim k^{-\gamma}$ with $\gamma=3$.
}
\label{pk}
\end{figure}
%
The BA model can be considered as a particular case 
of a model proposed by Simon \cite{simon} in 1955 to  
describe the scaling behaviour observed in 
distributions of words frequencies in texts, and in population 
figures of cities \cite{zipf}. 
The original Simon's model has been reformulated recently 
for networks growth in ref \cite{born}.
\\
In Fig.~\ref{pk} we report the degree distribution of a scale-free 
network obtained from the BA model (reported in black dots and indicated 
as $SF_{BA}$). 
We have constructed two networks, the first with 
$N=5000$, $K=10000$, and the second with $N=15000$, $K=75000$. 
In the first case the results reported are obtained as 
averages over 10 different realizations. While in the case $N=15000$ 
only one realization is sufficient to have a good 
statistics.

\subsection{Klemm-Egu\'{\i}luz (KE) scale-free networks}
\label{kesfn}
In this section  we introduce a  different class of 
scale-free networks with high clustering coefficient. We follow  the  
method developed by Klemm and Egu\'{\i}luz (KE) 
in Ref.\cite{klemm2}.
In the KE model, each node of the network is assigned a binary state variable 
and can be either in an active state or in 
a non-active state. 
Taking a completely connected network of $m$ active nodes 
as initial condition, the time-discrete dynamics of the 
KE model is based on the iteration of the following three steps: 

(1) {\it Addition of nodes}: 
A new node with $m$ edges is added to the network.  

(2) {\it Preferential attachment}: 
For each of the $m$ edges of the new node it is
decided with a probability $\mu$ whether the link connects 
to one of the active nodes or if it connects to a non-active 
node. In the latter case the random node is chosen according to the 
same rule of the BA model, the linear preferential attachment of 
eq. (\ref{pa}), i.e. the probability that node $i$ obtains a link is 
proportional to the node's degree: $\Pi(k_i)=k_i/{\sum_j k_j}$.  
The limit case $\mu = 1$ of the KE model is the BA model. 
The limit case $\mu = 0$ is a model with high clustering  
but large path length: in fact, as a function of the 
system size, $C$ quickly converges to a constant 
value, whereas $L$ increases linearly \cite{klemm1}. 

(3) {\it Activation and deactivation of nodes}. 
One of the $m$ active nodes is deactivated: the  
probability that node $i$ is chosen for deactivation is 
$\Pi^{deact}_i = k_i^{-1}/ \sum_l k_l^{-1}$. 
The new node is set in the active state.

The KE model generates scale-free networks with degree distribution 
$P(k)= 2 m^2 k^{-3}$ (for $k\ge m$) and average connectivity 
$<k> = \frac{2K}{N}=2 m$ \cite{klemm1}. Furthermore, by 
varying $\mu$ in the interval $[0,1]$ the model 
makes possible to study the cross-over 
between a case with high L and C 
(the model $\mu = 0$ has been studied previously 
in ref.\cite{klemm1}), and a case with small L and C 
($\mu = 1$ corresponds exactly to the BA model). 
Klemm and Egu\'{\i}luz have shown in Figure 1 of Ref.\cite{klemm2}
that a few ``long-range'' connections are sufficient to have a 
small-world transition: in fact, as soon as $\mu$ is different 
from zero, the average shortest path length $L$ 
drops rapidly approaching the minimum value of the BA model, 
while the clustering coefficient $C$ remains practically constant.
Thus the KE model with $\mu \neq 0$ and 
$\mu\ll1$ reproduces the three generic 
properties of real-world networks: 
power law degrees distribution, small $L$ and high $C$.   
In our simulations in the following of this paper, 
we have used the KE model with $\mu=0.1$. 
\\
In Fig.~\ref{pk} we report the degree distribution 
obtained for two different networks $N=5000$, $K=10000$ 
and $N=15000$, $K=75000$. 
Numerical simulations are shown both for the BA model 
(full circles) and the KE model (open squares). 
A good power-law behavior is obtained with an exponent 
$\gamma=3$ as expected.  The results for $N=5000$ 
are obtained as averages over 10 different realizations. 
While in the case 
$N=15000$ only one realization is sufficient to have a good 
statistics.

\subsection{Error and attack tolerance of BA scale-free networks}
\label{}
%
\begin{figure}
\begin{center}
\epsfig{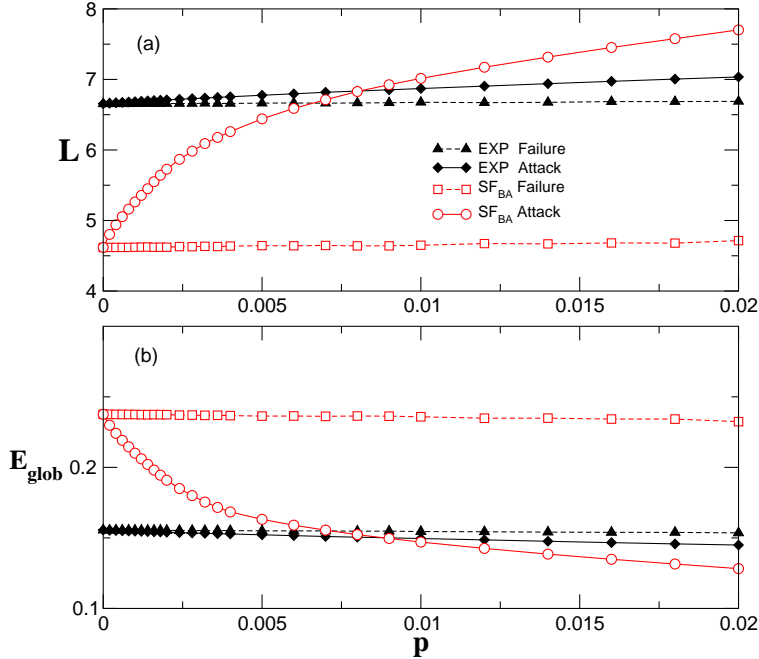}
\end{center}
\caption{
Resistance to failures and attacks: analysis of the global 
characteristics. BA scale-free graphs ($SF_{BA}$) are compared 
with random graphs (EXP). In both cases we start with 
two graphs with $N=5000$ nodes and $K=10000$ edges, and we 
remove a fraction $p$ of the nodes with two different 
prescriptions: failure and attack (see text).    
The correlation length $L$, in panel (a), 
and the global efficiency $\EGLOB$, in panel (b), 
are plotted as function of $p$.   
The results reported here and in all the following 
figures are averages over $10$ different realizations. 
}
\label{baglobalsmall}
\end{figure}
%
We are finally ready to address the problem of how the 
global and the local properties of a scale-free network are affected 
by the removal of some of the nodes. 
We consider first the class of scale-free networks generated 
by means of the BA model of Section \ref{basfn}.   
The malfunctioning of a node in general makes less 
efficient the communication between the remaining nodes, 
because it can eliminate some of the edges and consequentely 
some of the paths that contribute to the interconnectedness 
of the system. This will affect not only the global, but also
the local properties of the graph (though the latter have never been 
addressed in the literature before). 
As a starting point in our numerical experiments we consider 
a BA scale-free network with $N=5000$ nodes and $K=10000$ 
edges, corresponding to $<k>=4$. 
The error and attack tolerance of this network is compared to 
that of a random graph with the same number of nodes and edges. 
As previously mentioned, the $P(k)$ of a random graph is a 
Poisson distribution, a curve which for large $k$ 
decays exponentially and not as a power-law.
For this reason the random graph is indicated in the figures' 
captions as exponential graph (EXP). 
In removing the nodes, we use two different strategies. 
We can simulate an error in the system, as the {\it failure} 
of a node chosen at random among all the possible nodes. 
In alternative we can simulate an {\it attack} on the system 
by sorting the nodes in order of importance, according 
to their degree $k_i$, 
and then removing them one by one starting from the node 
with the highest degree. 
In fact an agent well informed about the whole structure 
of the network and wanting to deliberately 
damage the network will not target the
nodes randomly, but will preferentially attack  
the most connected nodes. 
Both for failures and attacks a fraction $p$ of the $N$ nodes is 
removed and the properties of the networks are studied computing 
the two quantities $L$ and $C$, or the two 
quantities $\EGLOB$ and $\ELOC$, as a function of $p$ 
(see Section \ref{}).  

\begin{figure}
\begin{center}
\epsfig{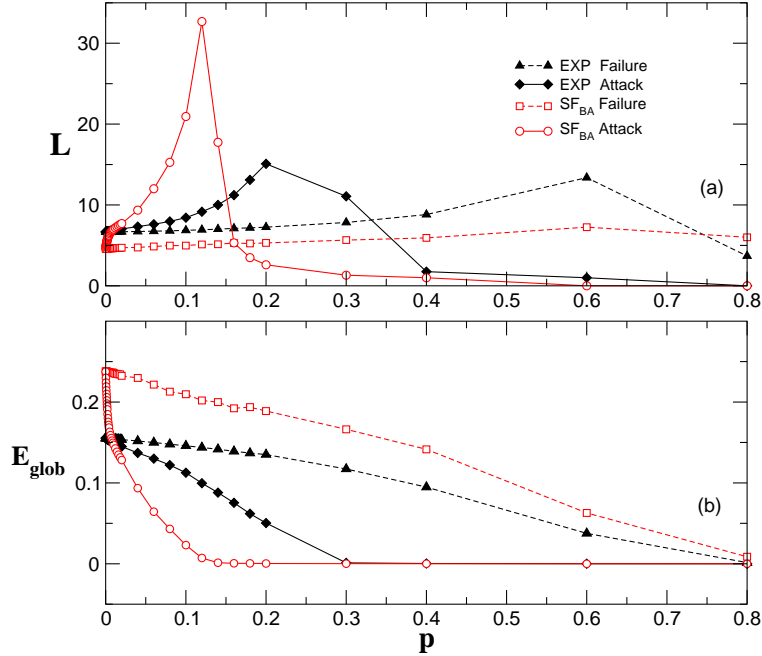}
\end{center}
\caption{Resistance to failures and attacks: analysis of the global 
characteristics. BA scale-free graphs are compared with random 
graphs. Same as in previous figure, but now the whole range of $p$ is 
considered.
}
\label{bagloballarge}
\end{figure}

{\bf Global properties.} 
In Fig.~\ref{baglobalsmall} and in  Fig.~\ref{bagloballarge}
we report $L$ and $\EGLOB$ as a function of the 
fraction $p$ of nodes removed. 
We first perform the same analysis of ref.\cite{barabasi3att} 
by studying the changes in the characteristic path length $L$. 
%
%
The scale-free graph considered initially has  $L \sim 4.6$ 
(on average, two generic nodes can be connected in less than 5 steps), 
a value lower than that of the random graph ($L \sim 6.7$).   
In the upper part of  Fig.~\ref{baglobalsmall}, we observe 
for the exponential network a slow 
monotonic increase of $L$ with $p$ (for $p \ll 1$), both for 
failures and for attacks. In practice there is no substantial 
difference whether the nodes are selected 
randomly or in decreasing order of connectivity. 
This behaviour is rooted in the homogeneity of the network: 
since all nodes have approximately the same number of links, 
they all contribute equally to the network characteristic 
path length, thus the removal of a generic node or the best connected 
one causes about the same amount of damage. 
On the other hand we oberve a drastically different 
behaviour for scale-free networks 
(the same observed in \cite{barabasi3att}): 
$L$ remains almost unchanged under an increasing level of errors, 
while it increases rapidly when the most connected nodes are 
eliminated. 
%
\begin{figure}
\begin{center}
\epsfig{figure=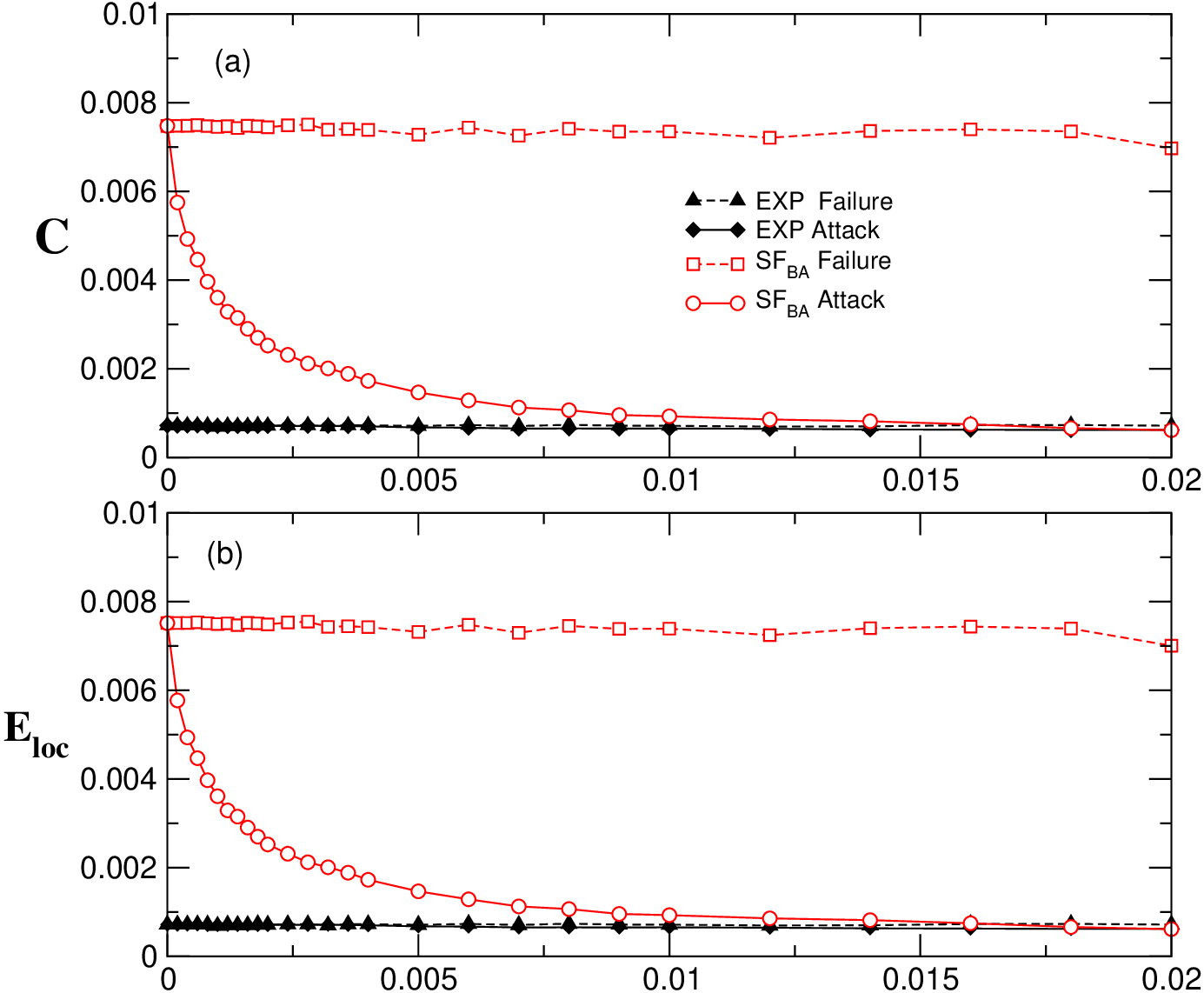,width=10truecm,angle=0}
\end{center}
\caption{Resistance to failures and attacks: analysis of the local  
characteristics. BA scale-free graphs are compared with random 
graphs. In both cases we consider two graphs with the same initial 
number of nodes $N=5000$ and edges $K=10000$.  
The clustering coefficient $C$, in panel (a), 
and the local efficiency $\ELOC$, in panel (b), 
are plotted as function of $p$, the fraction of nodes removed. 
}
\label{balocalsmall}
\end{figure}
%
For example, when $2\%$ of the nodes fails ($p=0.02$), 
the communication between the remaining nodes in the 
network is unaffected, while, when the $2\%$ of the most connected nodes 
is  removed, then $L$ almost doubles its original value.   
This robustness to failures and at the same time 
vulnerability to attacks is rooted in the inhomogeneity 
of the connectivity distribution $P(k)$: the
connectivity is maintained by a few highly connected nodes, 
whose removal drastically alters the network's topology, 
and decreases the ability of the remaining nodes to 
communicate with each other.
\\
In the following we show that 
this behavior can be better quantified by using 
$\EGLOB$, since the variable in formula \ref{globaleff} 
is normalized to the ideal case, obtained 
when all the $N(N-1)/2$ links are present in the graph. 
In the lower part of  Fig.~\ref{baglobalsmall} we observe that 
initially the scale-free graph has $\EGLOB=0.24$ and the random 
graph has $\EGLOB=0.15$, respectively $24\%$ and $15\%$ the efficiency 
of a completely connected graph. 
When $p=0.02$ and the nodes are removed under attack 
(i.e. according to their degree), the efficiency 
of the scale-free graph has rapidly decreased to $\EGLOB=0.12$: 
by attacking only a tiny fraction of nodes as the 
$2\%$, the scale-free network has already lost $50\%$ of its efficiency. 
Conversely the global efficiency of the scale-free graph does not vary a 
lot in the case of failures. 
The same thing happens for the exponential graph, where 
the communication between the remaining nodes of the
network is unaffected either from failures and from 
attacks. 
%
\begin{figure}
\begin{center}
\epsfig{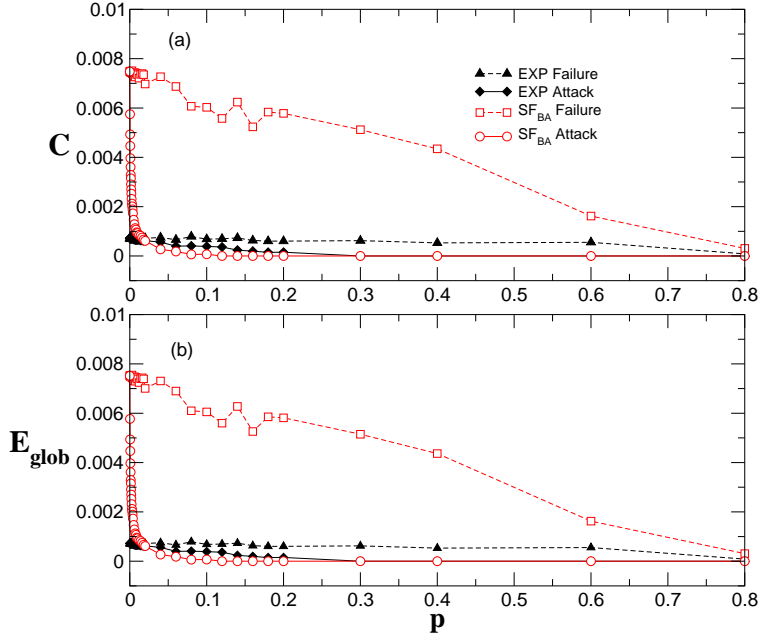}
\end{center}
\caption{
Resistance to failures and attacks: analysis of the local  
characteristics. BA scale-free graphs are compared with random 
graphs. Same as in previous figure, but now the whole range of 
$p$ is considered.
}
\label{balocallarge}
\end{figure}
%
\\
In so far we have only considered the removal of a small percentage 
of nodes. What happens now if we extend the analysis to larger values 
of $p$, even to values of the order of  1 ? 
In this case, it will become evident that the efficiency 
variable is a better quantity to study. 
In fact, for large values of $p$, we have to deal with the problem 
of the graph becoming unconnected. 
In the upper part of Fig.~\ref{bagloballarge}, we observe that 
$L$ reaches very high values when more and more 
nodes are removed. In practice, as explained in Section 
\ref{swn}, a straighforward application of the 
definition in formula (\ref{l}) would give $L=\infty$ 
for $p$ larger than a certain value $p^*$ for which the 
graph becomes unconnected. 
To avoid this divergence we have to limit the use of 
definition (\ref{l}) only to a part of the graph, 
the main connected component (as also done in \cite{barabasi3att}). 
In this way for different values of $p$ we compare graphs 
with different number of nodes, and this can 
give unrealistic results (see  Fig.~\ref{uncon}) 
as the maxima of $L$ observed in Fig.~\ref{bagloballarge}(a). 
See for example the BA scale-free network ($SF_{BA}$) 
under attacks: we have $L=30$ for 
$p=0.1$ and then a rapid drop to $L=4$ for $p=0.2$. 
This effect indicates that the network for $p=0.1$ starts 
to fragment into many unconnected small parts 
(each with more or less the same size) 
as evidenced from the cluster size distribution 
studied in Ref. \cite{barabasi3att}, 
but at the same time makes unfeasible 
the comparison of the connectivity properties of graphs with 
different $p$. In fact the misleading information we get 
from $L$ is that, by increasing $p$, i.e. by removing more nodes 
we can get a network with better connectivity (shorter $L$).  
In reality, when we want to compare graphs with $p$ varying in a 
wide range of values, it is better to use the  efficiency variable. 
In the lower part of Fig.~\ref{bagloballarge}, one can clearly 
see that, evaluating $\EGLOB$ as a function of $p$  
we get four monotonically decreasing curves, and 
we avoid the problem of the unphysical change of 
slope of $L$. 
Again we notice the rapid drop in the global efficiency of 
a scale-free network under attack: the removal of the $10\%$ 
of the nodes completely destroys the global efficiency that drops 
to values $\EGLOB \sim 0$. The removal of nodes by failure 
produces instead a slower decreases of $\EGLOB$ with $p$. 
When we compare these two curves with the two analog curves obtained 
for an exponential graph, we observe that in the case of a random 
graph the difference between failure and attack is less 
pronounced (though clearly visible on such a scale of $p$, while it 
was not visible in the short range $p$ scale used 
in Fig.~\ref{baglobalsmall}(b)) than in the case of the $SF_{BA}$ network. 
This means that, besides the sudden drop of $\EGLOB$ 
observed for $SF_{BA}$ under attack there are no other qualitative 
differences between scale-free and random graphs when their properties 
are compared on a large scale of $p$. 
\\  
The results we have reported in Fig.~\ref{baglobalsmall} 
and  Fig.~\ref{bagloballarge} are averages over $10$ different 
realizations. The average makes no important differences in the case of 
the global properties, although can be very important 
for the local quantities, which are in general affected by larger 
fluctuations. 

{\bf Local properties.} 
In Fig.~\ref{balocalsmall} and in  Fig.~\ref{balocallarge} 
we report $C$ and $\ELOC$ as a function of the nodes removed. 
We start, as before, with two networks, a BA scale-free and a 
random graph, with $N=5000$ nodes and $K=10000$ links. 
Of course both the networks considered have, by construction, 
a very small local clustering, as indicated by the small values 
of $C$ (0.007 for the BA scale-free network and less than 0.001 
for the random graph) or by the small values of 
$\ELOC$ (again 0.007 for the BA scale-free network and less 
than 0.001 for the random graph). 
The first thing to notice is that, in agreement with what said 
in Section \ref{swn}, the values of $C$ and $\ELOC$ are very similar. 
In fact we expect $C$ to be a reasonable 
approximation for $\ELOC$ when the 
subgraphs $\bf G_i$ 
of the neighbours of a generic node 
$i$ are composed by very small graphs \cite{lm2,lm4}.  
This is the case for both the random graph, and also the 
scale-free network of the BA model (things will be different for 
KE scale-free networks). 
Since the local clustering is very small we have large 
fluctuations among different realizations, and we must consider 
an average over different realizations to obtain stable results. 
The curves reported in Fig.~\ref{balocalsmall} and in 
Fig.~\ref{balocallarge} are averages over $10$ different 
realizations. 
Though the local clustering of the two networks is very small, 
we observe a rapid drop in the local efficiency 
(similarly to that observed for the global efficiency) 
of a scale-free network under attacks.

\subsection{Error and attack tolerance of KE scale-free networks}
\label{}
We now repeat the same analysis for the class of scale-free networks 
generated by the KE model, i.e. for networks with power law 
degree distribution and at the same time strong clustering. We can 
consider such networks as small-worlds with power-law degree 
distribution. 
We start by considering a KE scale-free $(SF_{KE})$
network with $N=5000$ nodes and $K=10000$ edges, 
generated by  the prescription of the KE model of 
Section~\ref{kesfn} with $\mu=0.1$.  
(such a scale-free network has also small-world properties, in fact it has 
$\EGLOB=0.12$ and $\ELOC=0.54$). 
As in the previous section we remove the nodes by using the 
two different strategies simulating failures or attacks, 
and we investigate how the properties of the network 
change by reporting as a function of $p$ 
the two quantities $L$ and $C$, or the two 
quantities $\EGLOB$ and $\ELOC$.  

\begin{figure}
\begin{center}
\epsfig{figure=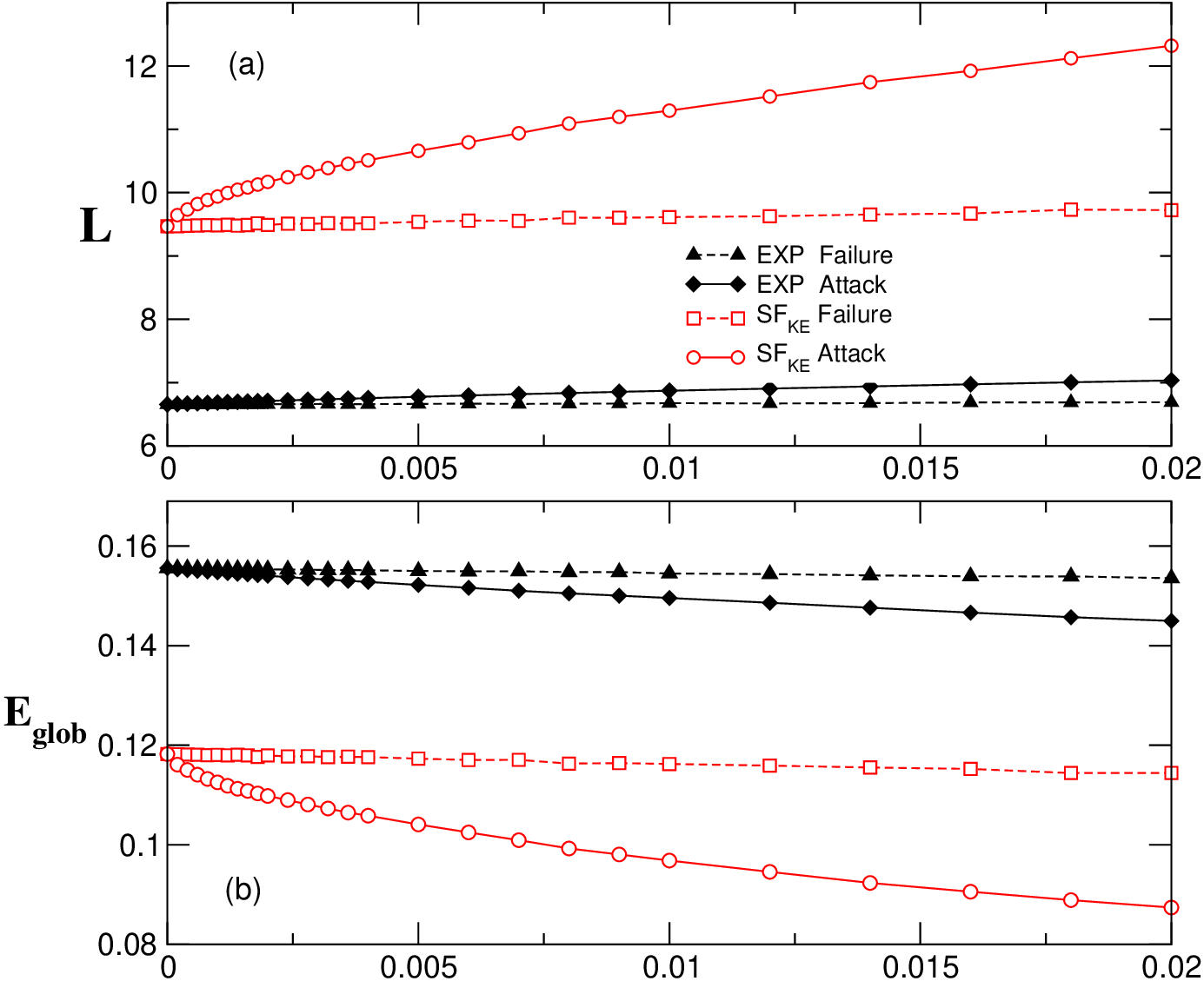,width=10truecm,angle=0}
\end{center}
\caption{
Resistance to failures and attacks: analysis of the global 
characteristics. KE scale-free graphs are compared with random 
graphs. In both cases we have two graphs with the same initial 
number of nodes $N=5000$ and edges $K=10000$. 
The correlation length $L$ and the global efficiency $\EGLOB$ 
are plotted as function of $p$, the fraction of nodes removed. 
}
\label{keglobalsmall}
\end{figure}

\begin{figure}
\begin{center}
\epsfig{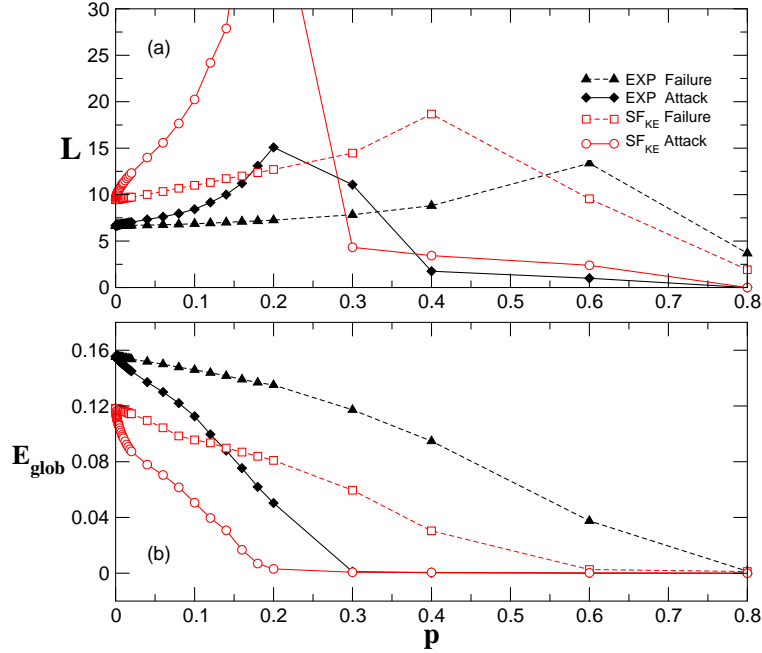}
\end{center}
\caption{
Resistance to failures and attacks: analysis of the global 
characteristics. KE scale-free graphs are compared with random 
graphs. Same as in previous figure, but now the whole range of $p$ is 
considered.
}
\label{kegloballarge}
\end{figure}

{\bf Global properties.}
In Fig.~\ref{keglobalsmall} and in  Fig.~\ref{kegloballarge}
we report $L$ and $\EGLOB$ as a function of the 
fraction $p$ of nodes removed.
%
The KE scale-free graph considered initially has now 
$L \sim 9.5$ (two generic nodes can be connected in an average 
of 10 steps). This value is higher than the value obtained 
for $SF_{BA}$ networks ($L \sim 4.6$), and also higher than that 
of random graphs ($L \sim 6.7$).  
This is of course the price to pay to have a strong local 
clustering: the increase in local connectivity is obtained 
at the expenses of the global connectivity. 
In any case, the results are similar to those  obtained for 
the BA scale-free networks, though the difference between 
scale-free and exponential network is less marked.  
In the upper part of  Fig.~\ref{keglobalsmall} we observe 
on one hand that the exponential network has a slow monotonic 
increase of $L$ with $p$ (for $p \ll 1$), 
both for failures and for attacks, and 
on the other hand that for scale-free networks 
$L$ remains almost unchanged under an increasing level of errors, 
while it increases rapidly when the most connected nodes are 
eliminated. 
%
%
In the lower part of figure Fig.~\ref{keglobalsmall} we see that the same 
behavior is confirmed when the global connectivity of the graph 
is expressed in terms of the efficiency $\EGLOB$:  
the initial efficiency  of the scale-free graph 
$\EGLOB=0.12$ ($12\%$ the efficiency 
of the completely connected graph) 
decreases to $\EGLOB=0.08$ by attacking the 
$2\%$ of the nodes (though this results is not as drastic as in the case 
of BA networks, compare with Fig.~\ref{baglobalsmall}). 
The global efficiency of the scale-free graph does not vary a 
lot in the case of failures. 
In Fig.~\ref{kegloballarge} we consider a larger range of 
values of $p$. 
From panel (a) we see again that the correct variable to evaluate 
is $\EGLOB$ and not $L$. In fact $L$ would give unphysical result 
as the presence of a spurious maximum when the network becomes 
unconnected.  
From the plot of $\EGLOB$ versus $p$ in Fig.~\ref{kegloballarge}(b) 
we observe that the KE scale-free and 
the exponential graph have a similar behavior as a function of $p$, 
when compared on the whole scale of $p$, apart from a different 
normalization factor, i.e. a different value at $p=0$. 
A qualitatively different behavior in the global properties 
of KE scale-free and exponential graphs is observed only for $p<0.02$ 
(compare Fig.~\ref{keglobalsmall} to  Fig.~\ref{kegloballarge}), i.e. 
only when a very small fraction of nodes is removed. 
%
\begin{figure}
\begin{center}
\epsfig{figure=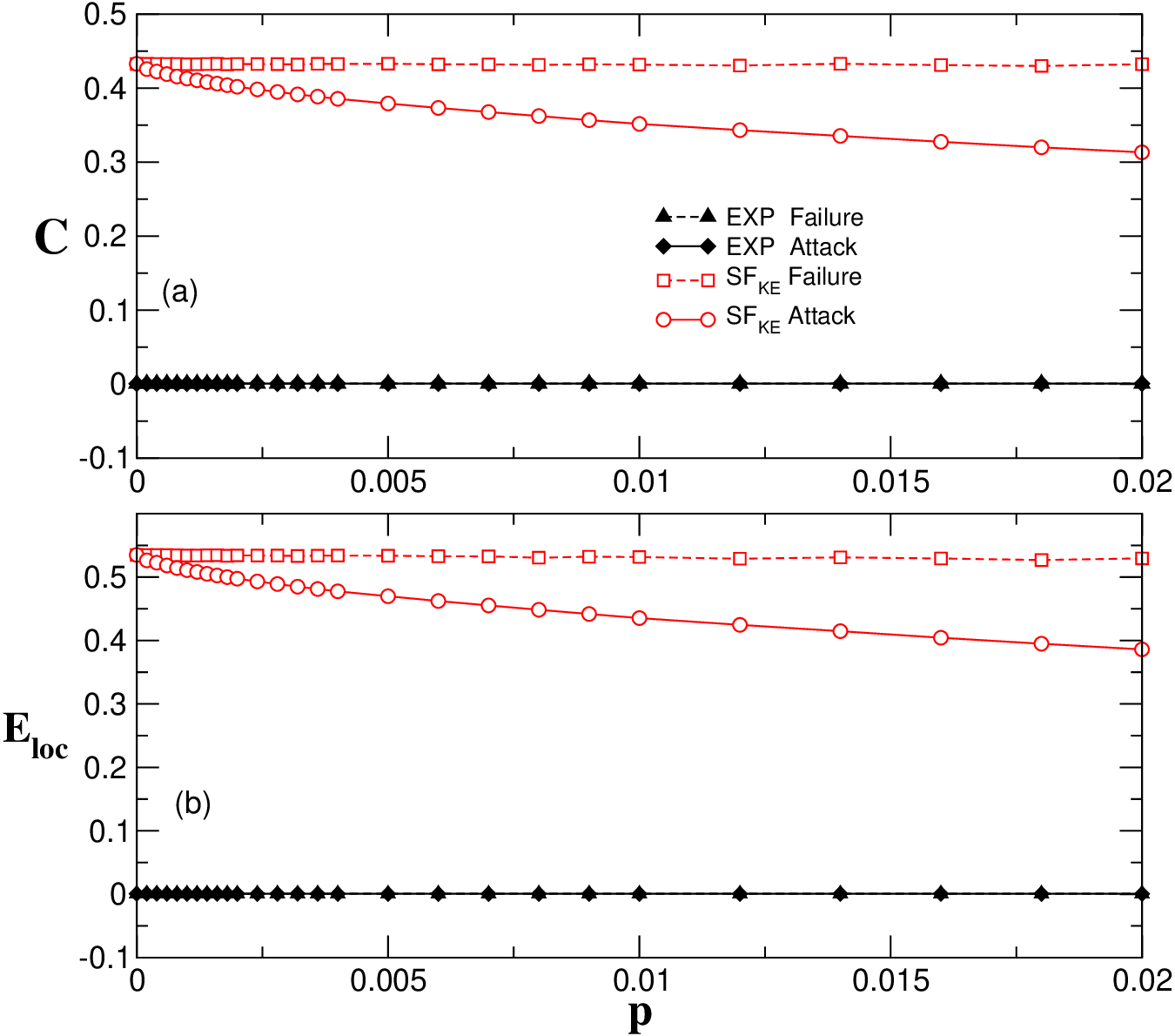,width=10truecm,angle=0}
\end{center}
\caption{Resistance to failures and attacks: analysis of the local  
characteristics. KE scale-free graphs are compared with random 
graphs. In both cases we have two graphs with the same initial 
number of nodes $N=5000$ and edges $K=10000$. 
The clustering coefficient $C$ and the local efficiency $\ELOC$ 
are plotted as function of $p$, the fraction of nodes removed. 
}
\label{kelocalsmall}
\end{figure}
%

\begin{figure}
\begin{center}
\epsfig{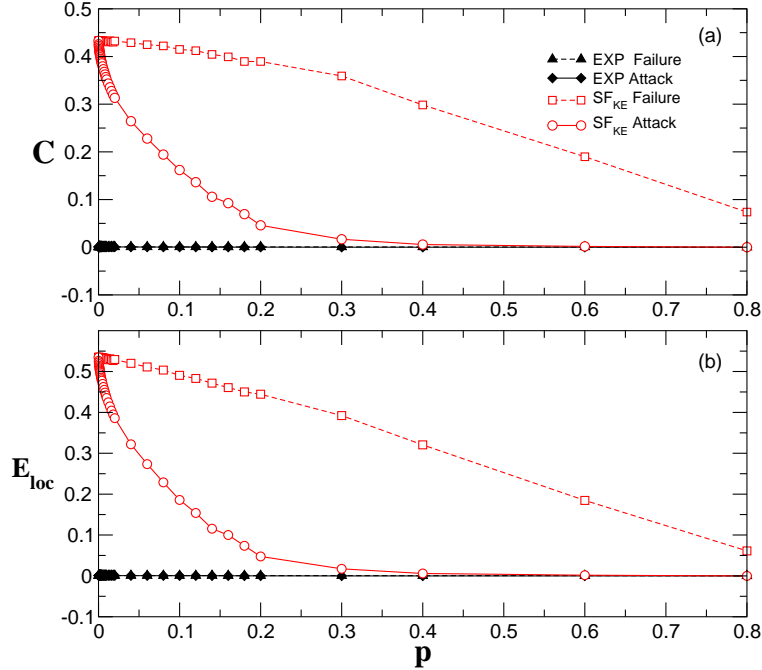}
\end{center}
\caption{
Resistance to failures and attacks: analysis of the local  
characteristics. KE scale-free graphs are compared with random 
graphs. Same as in previous figure, but now the whole range of 
$p$ is considered.
}
\label{kelocallarge}
\end{figure}

{\bf Local properties.} 
In Fig.~\ref{kelocalsmall} and in  Fig.~\ref{kelocallarge} 
we report $C$ and $\ELOC$ as a function of the nodes removed. 
We observe that the KE scale-free network has a good 
local connectivity expressed by a clustering coefficient 
$C=0.43$ and/or $\ELOC=0.54$ (meaning that the graph has 
$54\%$ of the local efficiency of the completely connected graph). 
Notice that, for KE scale-free networks the numerical values 
of $\ELOC$ and $C$ are not similar to each other, as they were in 
BA scale-free networks. 
In fact for $SF_{KE}$ networks the subgraph $\bf G_i$ of 
the neighbours of a generic node $i$ is not always 
a very small graph and therefore $C$ is not a good 
approximation of $\ELOC$ anymore \cite{lm2,lm4}. 
Though the numerical value of $C$ is different from that 
of $\ELOC$, the information we get from the behavior 
of these two quantities as a function of $p$ is similar.  
We observe, both in Fig.~\ref{kelocalsmall} and in 
Fig.~\ref{kelocallarge}, a rapid decrease in 
the local efficiency (and in the clustering coefficient $C$) 
of $SF_{KE}$ networks under attacks, 
while the local efficiency (and $C$) decreases much slower under 
failures. 
$\ELOC(p)$ and $C$ for random graphs, (the same curves were 
plotted in Fig.~\ref{balocalsmall} and Fig.~\ref{balocallarge} 
in larger scale), are here order of magnitude smaller than 
the values of the local efficiency of $SF_{KE}$ networks, 
and are practically indistinguishable from zero in the scale adopted.   
%
%

\section{ Conclusions}
\label{conclusions}
In this paper we have studied the effects of errors and 
attacks on the efficiency of scale-free networks. 
Two different kinds of scale-free networks have been  considered and compared 
to random graphs: 
scale-free networks with no local clustering produced by the Barabasi-Albert 
(BA) model, and scale-free networks with high clustering properties 
as in the model by Klemm and Egu\'{\i}luz (KE). 
By using as mathematical measures the global and the local 
efficiency, we have investigated the effects of errors 
and attacks both on the global and on the local properties of 
the network. 
We have found that both the global and the local efficiency 
of scale-free networks are unaffected by the failure of some of the nodes,  
i.e. when some (up to $2\%$) of the nodes are chosen at random and 
removed. On the other hand, at variance with random graphs, 
in scale-free networks the global and the local 
efficiency rapidly decrease when the nodes removed are those 
with higher connectivity $k_i$, i.e. scale-free networks 
are extremely sensitive to attacks. 
These properties are true both for BA networks and for KE networks, 
though KE networks have higher local efficiency but lower global 
efficiency than BA networks. 
We have also studied the effects of errors an attacks when a large   
number of nodes (even up to $80\%$ of the nodes of the network) 
are removed. On a such a larger scale of $p$ the difference between 
scale-free networks and random graph is less pronounced than in the 
smaller scale $p<0.02$. 
When a large number of nodes are removed, especially when 
the network become unconnected, the efficiency variable is 
definitely a better quantity than the characteristic path length 
$L$ to measure the response of the networks to external factors.

\bigskip
\noindent


\begin{thebibliography}{00}

%
\bibitem{yaneer} Y. Bar-Yam,
{\it Dynamics of Complex Systems} (Addison-Wesley, Reading Mass, 1997).

\bibitem{revstrogatz} S.H. Strogatz,
{\it Nature} {\bf 410}, 268 (2001).

\bibitem{revbarabasi} R. Albert and A.-L. Barab\'asi,
{\it Reviews of Modern Physics} {\bf 74}, 47 (2002). 

\bibitem{newman1} M.E.J. Newman, 
{\it J. Stat. Phys.} {\bf 101}, 819 (2000).

\bibitem{watts} D.J. Watts and S.H. Strogatz,
{\it Nature} {\bf 393}, 440 (1998).

\bibitem{milgram} S. Milgram,
{\it Psychol. Today}, {\bf 2}, 60 (1967).

\bibitem{lm2} V. Latora and   M. Marchiori
{\it Phys. Rev. Lett.} 87, 198701 (2001).

\bibitem{lm3} V. Latora and M. Marchiori, cond-mat/0202299, 
Proceedings of the International Conference 
``Horizons in Complex Systems'', 
Messina December 2001, to appear on Physica A.

\bibitem{lm4} V. Latora and M. Marchiori, 
cond-mat/0204089 and submitted to Phys. Rev. E. 

\bibitem{lm1} M. Marchiori and V. Latora, 
{\it Physica} {\bf A285}, 539 (2000).

\bibitem{barabasi0} R. Albert, H. Jeong, and A.-L. Barab\'asi,
{\it Nature\/} {\bf401}, 130 (1999).

\bibitem{barabasi1} A.-L. Barab\'asi and R. Albert,
{\it Science\/} {\bf286}, 509 (1999).

\bibitem{barabasi2}
A.-L. Barab\'asi, R. Albert and H. Jeong, 
{\it Physica} {\bf A272}, 173 (1999).

\bibitem{barabasi3att} R. Albert, H. Jeong, and A.-L. Barab\'asi,
{\it Nature\/} {\bf406}, 378 (2000); 
Correction {\it Nature\/} {\bf409}, 542 (2001). 

\bibitem{barabasi4met} H. Jeong, B. Tombor, R. Albert, 
Z.N. Oltvai and A.-L. Barab\'asi,
{\it Nature\/} {\bf407}, 651 (2000).  

\bibitem{barabasi5prot} H. Jeong , S.P.Mason, 
A.-L. Barab\'asi and Z.N. Oltvai, 
{\it Nature\/} {\bf411}, 41 (2001). 

\bibitem{bollobas} B. Bollob\'as,
{\it Random Graphs} (Academic, London, 1985).

\bibitem{klemm2}
K. Klemm, V.M. Egu\'{\i}luz, 
{\it Phys. Rev.\/} {\bf E65}, 057102 (2002). 

\bibitem{actors}
{\it The Internet Movie Database}, http://www.imdb.com

\bibitem{verme} T.B. Achacoso and W.S. Yamamoto,   
{\it AY's Neuroanatomy of C.\ elegans for Computation} (CRC Press,
Boca Raton, FL, 1992).

\bibitem{normalization} 
The formalism can be easily extended to the 
case of weighted networks \protect\cite{lm2,lm4}. 
Since in this paper we are interested in the study of 
unweighted networks we have directly presented the definition of 
the efficiency in the particular and simpler case of unweighted networks. 
In the general definition valid for weighted and unweighted networks 
a normalization factor has to be introduced to have: 
$0 \le \EGLOB({\bf G}) \le 1$ and 
$0 \le  \ELOC({\bf G}) \le 1$ (see refs. \cite{lm2,lm4}). 

\bibitem{wattsactors}
As done for example in ref. \protect\cite{watts} when the  
collaboration network of movie actors is studied, or in all the examples 
of \protect\cite{barabasi3att}. 

\bibitem{sta} F.L. Liljeros, C.R. Edling, N. Amaral,  
H.E. Stanley, and Y. Aberg, 
{\it Nature} {\bf 411}, 907 (2001).

\bibitem{ves} R. Pastor-Satorras and A. Vespignani, 
{\it Phys. Rev. Lett.} {\bf 86}, 3200 (2001). 

\bibitem{doye} J.P.K. Doye, cond-mat/0201430. 

\bibitem{pnas} L. A. N. Amaral, A. Scala, 
M. Barth\'el\'emy, and H.~E.~Stanley,
{\it Proc. Natl. Acad. Sci.} {\bf97}, 11149 (2000).

\bibitem{doro} S. Dorogovtsev, J. Mendes and A.N. Samukhin 
{\it Phys. Rev. Lett.\/}  {\bf85}, 4633 (2000). 
  
\bibitem{ley} P.L. Krapivsky, S. Redner, F. Leyvraz
{\it Phys. Rev. Lett.\/}  {\bf85}, 4629 (2000).

\bibitem{simon}
H.A. Simon, Biometrika {\bf 42}, 425 (1955).

\bibitem{zipf}
Zipf, G.K., {\em Human Behaviour and the Principle of Least Effort} 
(Addison-Wesley, Cambridge, Massachusetts, 1949).

\bibitem{born}
S. Bornholdt, H. Ebel, 
{\it Phys. Rev.\/} {\bf E64}, 035104(R) (2001).

\bibitem{klemm1}
K. Klemm, V.M. Egu\'{\i}luz, 
{\it Phys. Rev.\/} {\bf E65}, 036123 (2002).

\end{thebibliography}
\end{document}